\documentclass[fleqn,usenatbib,useAMS]{mnras}

\usepackage{graphicx}	
\usepackage{amsmath}	
\usepackage{amssymb}	
\usepackage{multicol}        
\usepackage{bm}		
\usepackage{pdflscape}	

\def \byzakaria\textcolor{ black}{ }

\newcommand\degree{^{\circ}}

\usepackage[T1]{fontenc}
\usepackage{ae,aecompl}

\usepackage{newtxtext,newtxmath}

\date{Last updated 2020 June 10; in original form 2013 September 5}

\pubyear{2022}

\title[Plerionic rectangular supernova remnants]{On the plerionic rectangular supernova remnants of static progenitors}

\author[D. M.-A.~Meyer et al.]
       {Meyer D. M.-A.$^1$, Meliani Z.$^2$, Velázquez P.~F.$^{3}$, 
       Pohl M.$^{1,4}$ and Torres D. F.$^{5,6,7}$  \\ 
       $^1$ Institut für Physik und Astronomie, Universität Potsdam, Karl-Liebknecht-Strasse 24/25, D-14476 Potsdam, Germany,\\
       E-mail: dmameyer.astro@gmail.com  \\ 
       $^2$ Laboratoire Univers et Théories, Observatoire de Paris, Université PSL, Université de Paris, CNRS, F-92190 Meudon, France\\
       $^3$ Instituto de Ciencias Nucleares, Universidad Nacional Aut\'onoma de M\'exico, CP 04510, Mexico City, Mexico \\
       $^4$ Deutsches Elektronen-Synchrotron DESY, Platanenallee 6, 15738 Zeuthen, Germany \\ 
       $^5$ Institute of Space Sciences (ICE, CSIC), Campus UAB, Carrer de Can Magrans s/n, 08193 Barcelona, Spain  \\ 
       $^6$ Institut d’Estudis Espacials de Catalunya (IEEC), Gran Capità 2-4, 08034 Barcelona, Spain \\
       $^7$ Institució Catalana de Recerca i Estudis Avançats (ICREA), 08010 Barcelona, Spain \\      
       }

\begin{document}
\label{firstpage}
\pagerange{\pageref{firstpage}--\pageref{lastpage}}
\maketitle

\begin{abstract}
Pulsar wind nebulae are a possible final stage of the circumstellar evolution of massive stars, where a fast-rotating, magnetised neutron star produces a powerful wind that interacts with the supernova ejecta. 
The shape of these so-called  plerionic supernova remnants is influenced by the distribution of circumstellar matter 
at the time of the explosion, itself impacted by the magnetic field of the ambient medium responsible for 
the expansion of the circumstellar bubble of the progenitor star. 
To understand the effects of magnetization on the circumstellar medium and resulting pulsar nebulae, we conduct 2D magneto-hydrodynamical simulations. 
Our models explore the impact of the interstellar medium magnetic field on the morphology of a supernova remnant and pulsar wind nebula that develop in the circumstellar medium of massive star progenitor in the warm phase of the Milky Way's interstellar medium.
Our simulations reveal that the jet-like structures formed on both sides perpendicularly to the equatorial plane of the pulsar, creating complex radio synthetic synchrotron emissions. 
This morphology is characterized by a rectangular-like remnant, which is typical of the circumstellar medium of massive stars in a magnetized medium, along with the appearance of a spinning top structure within the projected rectangle. We suggest that this mechanism may be partially responsible for the complex morphologies observed in pulsar wind nebulae that do not conform to the typical torus/jet or bow shock/tail shapes observed in most cases.
%
%
\end{abstract}



\section{Introduction}
\label{sect:intro}

Pulsar wind nebulae (PWN) are intriguing astrophysical phenomena powered by pulsars, rotating neutron stars with high magnetic field. 
%
PWN emerge from the intricate interplay between the pulsar's powerful wind and its surrounding environment, a phenomenon comprehensively elucidated in the review articles by \citet{Gaensler_Slane_2006ARA&A..44...17G, Kargaltsev_etal_2017JPlPh..83e6301K, Olmi_Bucciantini_2023PASA...40....7O}. These enigmatic nebulae manifest across the electromagnetic spectrum, predominantly characterized by synchrotron radiation and inverse Compton scattering, complemented by distinctive emission lines \citep{Reynolds_etal_2017SSRv..207..175R}. Their presence is evident from radio observations, exemplified by PWN G54.1+0.3 \citep{Driessen_etal_2018ApJ...860..133D}, to X-ray emissions typified by G11.2-0.3 \citep{Borkowski_etal_2016ApJ...819..160B}, and extends all the way to gamma-rays \citep{2006A&A...457..899A, Abdall_etal_2021ApJ...917....6A}, reaching PeV $\gamma$-ray energies (up to $10^{15}$ eV), as prominently demonstrated by the Crab Nebula  \citep{Lhaaso_Collaboration_2021Sci...373..425L}, see also \citet{acero_aph_2023}.

Pulsars represent the final evolutionary phase of massive stars that do not directly collapse into black holes. 
Understanding the physics of a pulsar and its interaction with the surrounding medium requires knowledge of various physical processes, including high-energy phenomena, fluid dynamics, general relativity, and nuclear physics, see, e.g., 
\cite{1999JPhG...25R.195W,Bucciantini_2011ASSP...21..473B,steiner_phr_411_2005,lasky_pasa_32_2015,2015PhRvD..92h4018P,2016PhRvD..94b4053P,pnigouras_phrvd_2019}. 
%
Pulsars have very powerful magnetospheres with strong magnetic fields on the order of kilogauss (kG), which play a crucial role in the evolution 
see e.g.,
\cite{mestel_mnras_217_1985}. 
The rotating magnetospheres extract energy from the pulsar and generate a powerful wind, see, e.g., \cite{petri_aa_666a_2022}. 
The interaction of the pulsar wind with the ambient medium produces the so-called pulsar nebulae, which can be located inside or outside the supernova remnant of the progenitor star, depending on whether the supernova explosion had kicked off the pulsar.
Various observations have triggered investigations into such phenomena, see, amongst others, the studies of \citet{Pavan_etal_2016A&A...591A..91P, Kargaltsev_etal_2017JPlPh..83e6301K,deVries_Romani_2020ApJ...896L...7D,igoshev_mnras_494_2020,vries_apj_908_2021}.

A particular class of supernova remnants containing a pulsar exhibit a succession of structured shocks powered by the pulsar's magnetic wind, producing multi-wavelength polarized non-thermal emission. 
Examples of such plerions include the Crab Nebula~\citep{Hester_2008ARA&A..46..127H}, as well as the Geminga pulsar residing within the Vela supernova remnant~\citep{bock_aj_1998, popov_2019}. Additionally, one can observe youthful supernova remnants hosting both a pulsar and a pulsar-wind nebula, such as B0540-693 \citep{Williams_etal_2008ApJ...687.1054W} and G11.2-0.3 \citep{Borkowski_etal_2016ApJ...819..160B}.

The modelling of PWN has been a long-standing challenge for several reasons. 
First, the physics involved in PWN is inherently complex, involving the interaction between the pulsar's relativistic wind and the surrounding medium. 
This requires a multi-disciplinary approach. 
Second, the environment in which the pulsar wind is launched is often structured,
as it depends on the supernova remnant's properties, the progenitor star's circumstellar medium, or the interstellar medium in which the pulsar resides. 
The 
properties of the surrounding medium can significantly affect the dynamics and emission of the PWN.
These factors together make the modelling of PWN a complex and multi-faceted problem, requiring sophisticated theoretical models and numerical simulations to understand the physics at play fully. 
%
%
%
%
%
%
The Crab Nebula stands out as a prominent example of a plerion. Extensive research, both theoretical~\citep{kennel_apj_283_1984,coroniti_apj_349_1990,begelman_apj_397_1992,begelman_apj_493_1998} and numerical, has been dedicated to studying PWN like the Crab Nebula. These investigations encompass relativistic axisymmetric 2D simulations  \citep[e.g.,][]{komissarov_mnras_344L_2003,komissarov_mnras_349_2004, komissarov_mnras_367_2006, Del_Zanna_etal_2006A&A...453..621D, Camus_etal_2009MNRAS.400.1241C, komissarov_mnras_414_2011, Olmi_etal_2014MNRAS.438.1518O} as well as relativistic 3D simulations 
\citep[e.g.,][]{Mignone_mnras_436_2013, Porth_etal_2014MNRAS.438..278P, Olmi_etla_2016JPlPh..82f6301O}.

\begin{table*}
	\centering
	\caption{
	List of models in this study. All simulations assume a rotating static massive star
	of mass $M_{\star}$ at solar metallicity, in a medium of number density $n_{\rm ISM}$ and 
    organised magnetic field strength $B_{\rm ISM}$. The initial rotation rate of the central 
    massive star is $\Omega_{\star}/\Omega_{\rm K}=0.1$. 
	}
	\begin{tabular}{lcccr}
	\hline
	${\rm {Model}}$         &  $M_{\star}$ ($\rm  M_{\odot}$)  &    $n_{\rm ISM}$   ($\rm  cm^{-1}$)     
    &   $B_{\rm ISM}$ ($\mu\, \rm  G$)  
                         	&    Description     \\ 
	\hline   
	\normalfont{Run-35-HD-0.79-PWN}          &  $35$  &   $0.79$     &  0 	&  $35$$-$$\rm  M_{\odot}$$-$induced PWN in 
    unmagnetised warm ISM         \\
	\normalfont{Run-35-MHD-0.79-PWN}         &  $35$  &   $0.79$     &  7 	&  $35$$-$$\rm  M_{\odot}$$-$induced PWN in 
    magnetised warm ISM         \\
	\hline    
	\end{tabular}
\label{tab:table1}
\end{table*}

Moreover, the Crab Nebula is pivotal in advancing our understanding of pulsar physics and their interactions with supernova remnants. Notably, the dynamics and morphology of pulsar wind nebulae experience significant transformations as they expand within a supernova remnant. 
%
This influence can be even more pronounced when the pulsar receives a kick during a supernova explosion, as observed in the case of PWN CTB 87 \citep{Matheson_etal_2013ApJ...774...33M}. 
Extensive studies have been conducted on the interaction between PWN and supernova remnants, focusing on no-moving pulsars in 1D and 2D scenarios \citep{swaluw_aa_380_2001,blondin_apj_845_2017}, including a mock complex surrounding the neutron star \citep{swaluw_aa_404_2003,blondin_apj_563_2001}. 
%

These studies have also been extended to moving pulsars inside the supernova ejecta, revealing the development of strong bow-shocks between the pulsar wind and supernova remnant  \citep{swaluw_aa_397_2003,swaluw_aa_420_2004,temim_apj_851_2017,kolb_apj_844_2017,temim_apj_932_2022}
 and the study was extended using relativistic MHD \citep{bucciantini_aa_422_2004}.
In some cases, the strong interaction between the PWN 
and the reverse shock of the supernova remnant can result in compression. 
This interaction phase is referred to as reverberation, during and after which the morphology, spectrum, and dynamics of PWN could undergo significant changes, see the recent studies by
\cite{Torres_Lin:2018,Bandiera_mnras_499_2020,Bandiera_mnras_165_2023,2023_mnras_2839_2023}.

In later phases, as the moving pul
sar leaves the supernova remnant and begins to interact with the interstellar medium, a 
bow shock nebula forms around the runaway pulsar. This intriguing phenomenon has been extensively studied analytically \citep{Bucciantini_aa_375_2001} and through numerical simulations in two dimensions with relativistic considerations 
 \citep{bucciantini_mnras_478_2018}. 
Furthermore, research into this phenomenon has delved into three-dimensional simulations, 
encompassing non-relativistic pulsar winds \citep{Toropina_etal_2019MNRAS.484.1475T}
and relativistic pulsar winds
\citep{barkov_mnras_484_2019, barkov_2019MNRAS.485.2041B, barkov_mnras_497_2020,Olmi_Bucciantini_2019MNRAS.488.5690O}.
%
%
%
Despite the high complexity of these simulations and the numerous questions they leave unanswered~\citep{Olmi_Bucciantini_2023PASA...40....7O}, all of these studies still neglect the effects of the circumstellar medium of the defunct star in which the supernova remnant and the PWN expand during the initial phases.
%

%
The circumstellar medium of a defunct star is formed through the interaction between the star's wind and luminosity with the surrounding interstellar medium (ISM). 
The shape and properties of the circumstellar medium depend on various factors, including the evolution of the star, such as its mass, age, and stage of evolution, as well as the characteristics of the surrounding ISM, such as density, temperature, and magnetic field \citet{vanmarle_2015,meyer_515_mnras_2022}. 
%
%
In the context of massive stars, the circumstellar medium undergoes successive structural changes. During its early life, it forms an accretion disc~\citep{liu_apj_904_2020,meyer_mnras_517_2022,elbakyan_mnras_518_2023,burns_NatAs_2023}. In its main sequence phase, it expands into a wind bubble~\citep{weaver_apj_218_1977,gull_apj_230_1979,wilkin_459_apj_1996}. Later on, it evolves into expanding shells~\citep{stock_mnras_409_2010,cox_aa_537_2012,decin_asr_50_2012,2ddddddd}. If a supernova explosion occurs, it leaves behind an expanding remnant shell~\citep{aschenbach_aa_341_1999,yusefzadeh_apj_585_2003,katsuda_apj_863_2018,arias_aa_622_2019,2019arXiv190908947C,2019arXiv190906131D}.
Once the pulsar emits a relativistic and powerful wind, it initially interacts with the surrounding supernova ejecta~\citep{cox_apj_524_1999,sun_apj_511_1999,crawford_apj_554_2001,Olmi_Bucciantini_2023PASA...40....7O}. As the PWN 
passes through the supernova ejecta, it subsequently interacts with the circumstellar medium of the defunct star. The distribution of ejecta, stellar wind, and ISM gas acts as a matrix that channels the expansion of the pulsar wind~\citep{kolb_apj_844_2017,temim_apj_932_2022}. 
This is particularly important when the supernova progenitor is a runaway star, as the bow shock created by its surrounding stellar wind can influence the subsequent evolution and emission of the supernova ejecta and the 
PWN~\citep{meliani_mnras_515_2022}.

%
This study aims to investigate how a magnetised ambient medium influences the dynamics, morphologies, and emission properties of PWN with static massive star progenitors.
The multi-dimensional magnetohydrodynamical (MHD) simulations conducted by \citet{vanmarle_584_aa_2015} have revealed that the circumstellar medium of high-mass stars is significantly influenced by the organized magnetic field of its ambient medium. This finding has profound implications on the understanding of stellar wind bubbles around massive stars, as previously studied by \citet{freyer_apj_594_2003, dwarkadas_apj_630_2005, freyer_apj_638_2006, dwarkadas_apj_667_2007}.
 The presence of a magnetic field can cause expanding stellar wind bubbles to become elongated and adopt an oblong morphology along the direction of the magnetic field lines. Our previous work \citep{meyer_515_mnras_2022} has shown that such asymmetric pre-supernova environments can result in a peculiar reflection of the supernova shock wave, forming rectangular-shaped remnants like Puppis A. In this study, we further investigate the effects of the reflection of the supernova blastwave in asymmetric, magnetized wind bubbles that are generated by a static, rotating star in the warm phase of the Galactic plane and how it may impact the evolution of plerionic pulsar wind nebulae.

%
The paper is structured as follows. In Section~\ref{sect:method}, we present the modelling methods used in this study. This includes the description of the numerical simulations of pulsar wind nebulae, which are detailed in Section~\ref{sect:results}. We then discuss the outcomes of our study in Section~\ref{sect:discussion}, and present our conclusions in Section~\ref{sect:conclusion}.


\section{Method}
\label{sect:method}

In this section, we will provide a comprehensive review of the numerical setup used in this study to generate models of PWN from static massive stars. 
We will summarize the initial conditions, including both the initial and boundary conditions, in the following paragraphs. Additionally, we will describe the numerical methods employed in the simulations.

\subsection{Initial conditions and boundaries conditions}
\label{initial_conditions}

 This paper presents models that simulate the interaction between a star's wind and ejecta at all phases of its evolution with the warm ISM in the Milky Way galaxy. 
 The total number density of the ISM is taken to be $n_{\rm ISM}=0.79\, \rm cm^{-3}$, while the magnetic field of the ISM is uniform and structured, with a strength of $B_{\rm ISM}=7\, {\rm \mu G}$. 
 In these models, we assume that the ionized gas has a temperature of $8000\, \rm K$ (table~\ref{tab:table1}). 
 The ambient medium is in equilibrium between the photoheating provided by the reionizing gas around the star, as described in \citet{osterbrock_1989} and \citet{hummer_mnras_268_1994}, and the radiative losses from optically-thin cooling processes, as outlined in \citet{wolfire_apj_587_2003}.
%
 The cooling law used in this study is based on the work of~\citet{wiersma_mnras_393_2009}, which is suitable for a solar metallicity environment~\citep{asplund_araa_47_2009}. 
 The cooling law accounts for hydrogen and helium as the primary coolants at temperatures $T<10^{6}\, \rm K$, and various metals' emission lines at temperatures $T \ge 10^{6}\, \rm K$. 
 The cooling curve is further enhanced with [O{\sc iii}] $\lambda , 5007$ collisionally excited forbidden lines, as described in \citet{asplund_araa_47_2009} and \citet{henney_mnras_398_2009}.

This paper presents a model that captures the evolution of the circumstellar medium surrounding a static massive star with an initial mass of $35\, \rm M_{\odot}$ at the zero-age main sequence. 
The star is considered to be rotating with an angular velocity ratio of $\Omega_{\star}/\Omega_{\rm K}=0.1$, where $\Omega_{\star}$ represents the star's initial angular frequency and $\Omega_{\rm K}$ is its equatorial Keplerian angular velocity.
Consequently, the equatorial velocity of the star can be expressed as,
\begin{equation}    
     v_{\rm rot}(t) = \Omega_{\star}(t) R_{\star}(t).
\end{equation}
Here, $R_{\star}(t)$ denotes the stellar radius, and the time-dependence signifies the variation in surface properties throughout the star's entire lifespan.
%
The model tracks the complete evolution of the circumstellar medium surrounding the static star, ranging from the onset of the zero-age main sequence to the pre-supernova phase. This comprehensive approach encompasses various stages, including the main sequence, red supergiant, and final Wolf-Rayet phase.

Regarding the stellar wind throughout the evolution phase, we assume that the stellar wind maintains spherical symmetry throughout the entire lifespan of the supernova progenitor, with the axis of rotation of the rotating star aligned with the axis of symmetry of the domain. 
To determine the wind's characteristics, we use the one-dimensional stellar evolution model provided by the {\sc geneva} library, as described in~\citet{ekstroem_aa_537_2012}\footnote{https://www.unige.ch/sciences/astro/evolution/en/database/syclist/}. 
Specifically, we extract the mass-loss rate $\dot{M}(t)$ and the effective temperature $T_{\rm eff}(t)$ of the star at each stage of evolution from this database, and derive from it the wind density, 
\begin{equation}
	\rho_{w}(r,t) = \frac{ \dot{M}(t) }{ 4\pi r^{2} v_{\rm w}(t) }, 
    \label{eq:wind}
\end{equation}
In this equation, $r$ represents the radial distance from the star, and $\dot{M}(t)$ corresponds to the mass loss rate of the star at time $t$.

The terminal velocity of the stellar wind, denoted as $v_{\rm w}(t)$, is calculated based on the escape velocity $v_{\rm esc}(t)$. 
The escape velocity depends on the star's effective temperature $T_{\rm eff}$ and is determined using the conversion law,  
\begin{equation}    
     v_{\rm w}(t) = \sqrt{ \beta(T) } v_{\rm esc}(t) 
     = \sqrt{ \beta(T)  \frac{  2GM_{\star}(t) }{ R_{\star}(t)} },
\end{equation}
where $G$ represents the gravitational constant, and $\beta(T)$ is a the normalisation factor introduced by \cite{eldridge_mnras_367_2006}.

We adopt the time-dependent evolution of the surface magnetic field $B_{\star}$ of the supernova progenitor as derived in~\citet{Meyer_etal_2023MNRAS.521.5354M} where the magnetic field strength at the surface of the star are scaled to that of the Sun, as described in ~\citet{scherer_mnras_493_2020,herbst_apj_897_2020,baalmann_aa_634_2020,baalmann_aa_650_2021,meyer_mnras_506_2021}. 
Specifically, we assume a magnetic field strength at star surface of $B_{\star}=500\, \rm G$ during the main-sequence phase~\citep{fossati_aa_574_2015,castro_aa_581_2015,  przybilla_aa_587_2016,castro_aa_597_2017}, to a Betelgeuse-like field of $B_{\star}=0.2\, \rm G$ for 
the red supergiant phase~\citep{vlemmings_aa_394_2002,vlemmings_aa_434_2005, 
kervella_aa_609_2018}  and $B_{\star}=100\, \rm G$ during the Wolf-Rayet phase ~\citep{meyer_mnras_507_2021}.  
Concerning the stellar magnetic field structure, we utilize a Parker spiral made of 
a radial component, 
\begin{equation}
	B_{\rm r}(r,t) = B_{\star}(t) \Big( \frac{R_{\star}(t)}{r} \Big)^{2},
    \label{eq:Br}
\end{equation}
and a toroidal component,
\begin{equation}
	B_{\phi}(r,t) = B_{\rm r}(r,t) 
	\left( \frac{ v_{\phi}(\theta,t) }{ v_{\rm w}(t) } \right) 
	\left( \frac{ r }{ R_{\star}(t) }-1 \right),
    \label{eq:Bphi}
\end{equation}
respectively, with, 
\begin{equation}
	v_{\phi}(\theta,t) = v_{\rm rot}(t) \sin( \theta ),
\label{eq:Vphi}
\end{equation}
being the latitude-dependent surface velocity of the rotating massive star~\citep{parker_paj_128_1958,weber_apj_148_1967,pogolerov_aa_321_1997,
pogolerov_aa_354_2000,chevalier_apj_421_1994,rozyczka_apj_469_1996}. 
%

%
At the end of a star's evolution, it enters the supernova phase, during which we model the expanding supernova ejecta as a spherically symmetric distribution within a radius of $r_{\rm max}$. 
The ejecta has a total energy of $E_{\rm SN}=10^{51}$ erg and a mass of $M_{\rm SN}$, which takes into account the star's mass loss throughout its entire evolution until the immediate pre-supernova time $t_{\rm psn}$, as well as the mass $M_{\rm NS}$ of the neutron star that forms at the centre. Specifically, we set, 
\begin{equation}
M_{\rm sn}= M_{\star} - \int_{0}^{t_{\rm psn}} \dot{M}(t) dt - M_{\rm NS}=10.12 M_{\odot}, 
\end{equation}
with $t_{\rm psn}$ and $M_{\rm NS}=1.4 \, \rm M_{\odot}$ \citep{das_aa_661_2022}.

In our study, we adopt a density and velocity profile for the freely 
expanding supernova ejecta based on the work by \cite{truelove_apjs_120_1999}. 
This profile consists of two distinct regions \citep{Bandiera:2021}. 
The first region is a uniform density core extending from 0 to $r_{\rm core}$, where $r_{\rm core}$ represents the core radius. In this region, the density decreases with time following a power-law relationship of $t^{-3}$, where $t$ denotes the time after the explosion.
The second region is the outer edge, extending from $r_{\rm core}$ to $r_{\rm max}$, where $r_{\rm max}$ corresponds to the maximum radius. In this region, the density decreases steeply with radius, following a power-law relationship of $\rho \propto r^{-n}$, with the exponent $n$ set to 11. Additionally, the density in the outer edge region decreases with time as $t^{-(3+n)}$.
These density profiles can be expressed as follows:
\begin{equation}
   \rho_{\rm core} =  \frac{1}{ 4 \pi n } \frac{ \left(10 E_{\rm sn}^{n-5}\right)^{-3/2}
 }{  \left(3 M_{\rm sn}^{n-3}\right)^{-5/2}  } \frac{ 1}{t_{\rm max}^{3} },
   \label{sn:density_1}
\end{equation}
and, 
\begin{equation}
   \rho_{\rm max}(r) =  \frac{1}{ 4 \pi n } \frac{ \left(10 E_{\rm
sn}^{n-5}\right)^{(n-3)/2}  }{  \left(3 M_{\rm sn}^{n-3}\right)^{(n-5)/2}  } \frac{ 1}{t_{\rm max}^{3} } 
\bigg(\frac{r}{t_{\rm max}}\bigg)^{-n},
   \label{sn:density_2}
\end{equation}
respectively. These density profiles are commonly used for core-collapse supernovae \citep{chevalier_apj_258_1982}.

For the velocity, we utilize a homologous radial profile for the supernova ejecta, given by $v=t/r$, across all regions from 0 to $r_{\rm max}$ at time $t_{\rm max}$. The characteristics of the supernova ejecta profile are computed following the methodology outlined in  \cite{truelove_apjs_120_1999} and \cite{whalen_apj_682_2008}.

The velocity at the core radius, denoted as $v_{\rm core}(r_{\rm core})$, is determined as, 
\begin{equation}
v_{\rm core}(r_{\rm core}) = \left(\frac{ 10(n-5)E_{\rm sn} }{ 3(n-3)M_{\rm sn} }\right)^{1/2},
\label{sn:vcore}
\end{equation}
where $E_{\rm sn}$ represents the total energy of the supernova ejecta and $M_{\rm sn}$ is the 
total mass of the ejecta. This equation ensures conservation of both mass and energy in the 
supernova ejecta. The maximum speed, denoted as $v_{\rm max}$, is set to:
\begin{equation}
v_{\rm max}= \frac{r_{\rm max}}{t_{\rm max}}=3\times 10^{4}\, \rm km\, \rm s^{-1},
\end{equation}
This choice of $v_{\rm max}$ maintains the conservation of total mass and energy in the supernova 
ejecta \citep{vanveelen_aa_50_2009}.

As the supernova ejecta are expelled, we set a radial pulsar's wind that emanates from the centre, as described by \citet{meyer_515_mnras_2022}. 
This wind has a total power that is assumed to evolve over time, $t$, according to the following equation:
\begin{equation}
\dot{E}_{\rm psw}=\dot{E}_{\rm psw,0}\left(1+\frac{t}{\tau_{0}}\right)^{-\frac{n+1}{n-1}}. 
\end{equation}
with $\tau_{\rm o}$ is the initial spin-down of the pulsar, defined as,
\begin{equation}
\tau_{\rm o} = \frac{ P_{\rm o}  }{   (n-1)\dot{P}{\rm o}  }, 
\end{equation}
where the initial spin period of the pulsar is set to $P_{\rm o}=0.3\, \rm s$, and its time derivative is set to $\dot{P}_{\rm o}=10^{-17}\, \rm s\, \rm s^{-1}$.
%
The braking index is assumed to be $n=3$, which corresponds to magnetic dipole spin-down, as outlined in \citet{Pacini_1967Natur.216..567P}.

Furthermore, we assume that the pulsar's wind maintains a constant speed of $v_{\rm psw}=10^{-2}c$, where $c$ denotes the speed of light in vacuum. It's important to acknowledge that this speed is significantly lower than the realistic pulsar wind speeds, which can reach $c$, corresponding to a Lorentz factor of $10^6$ \citep[as demonstrated in][]{kennel_apj_283_1984}. This decision to employ a reduced pulsar wind speed can lead to noticeable alterations in the properties of its termination shock. These changes encompass compression rates, speeds, and, subsequently, shock positions and influence the development of associated instabilities.
It is crucial to emphasize that our paper's primary objective is to replicate the overall evolution of the PWN accurately. This evolution is predominantly governed by the wind's momentum flux \citep{Wilkin_1996ApJ...459L..31W}.
In terms of magnetization, we have opted for a low value of $\sigma=10^{-3}$ in this study, a choice in line with descriptions found in \citet{Rees_Gunn_1974MNRAS.167....1R}, \citet{kennel_apj_283_1984}, \citet{2017hsn_book_2159S}, \citet{begelman_apj_397_1992} and \citet{TORRES_et_al_201431}. 
This magnetization value implies that a significant portion of the magnetic field is 
converted into kinetic energy.
%
It is worth noting that recent multi-dimensional simulations have demonstrated that larger magnetization values, such as $\sigma=0.01$ in  2D \citep[e.g.,][]{komissarov_mnras_344L_2003,komissarov_mnras_349_2004,zanna_aa_421_2004,Del_Zanna_etal_2006A&A...453..621D} and even $\sigma>1$ in 3D \citep{Porth_etal_2014MNRAS.438..278P, barkov_mnras_484_2019}, can accurately reproduce the features of termination shocks of PWN. 
%

However, it is essential to recognize that the value of pulsar wind magnetization remains a topic of debate, as it significantly influences PWN termination shock strength and, consequently, particle acceleration. 
%
%
Moreover, the magnetization of the equatorial wind zone may decrease, leading to lower magnetization values due to the annihilation of equatorial wind magnetic stripes \citep{Coroniti_2017ApJ...850..184C}. By selecting such a low magnetization, as in \citet{bucciantini_aa_422_2004}, the Pulsar Wind Nebula tends to expand more in the equatorial plane, resulting in a stronger termination shock. 

%
Furthermore, Komissarov and Lyubarsky in 2003-2004 and Del Zanna et al. 2004, showed that 
the properties of the inner nebula can only be recovered with 2D simulations if the 
injected magnetization is larger than ~ 0.01. 
%
%

The magnetic field is assumed to have only a toroidal component. The total kinetic energy, magnetic field strength, and kinetic energy are functions of the radial distance $r$ and polar angle $\theta$, as described in \citet{komissarov_mnras_349_2004}.
\begin{eqnarray}
\dot{E}_{\rm total}=\frac{\dot{E}_{\rm psw}}{r^2}\left(\sin^2{\theta}+1/\sigma\right),\\
B=\sqrt{\frac{4\pi}{c}}\frac{1}{r}\sin{\theta} \left(1-2\theta/\pi\right),\\
\dot{E}_{\rm kinetic}=\dot{E}_{\rm total}-\frac{B^2}{4\pi}\,c\,.
 \end{eqnarray}

Our choice of a spherically symmetric supernova explosion allows us to assume that the neutron star is at rest at the location of the explosion and neglects any potential kick velocity resulting from asymmetries in the explosion.

\subsection{Numerical methods}
\label{numerical_methods}

To investigate the evolution of the PWN within the circumstellar medium of its static progenitor star that is surrounded by a magnetized external medium, we follow the strategy we used in \citet{meyer_mnras_450_2015, meyer_mnras_493_2020} and that we extended after to PWN in \citet{meliani_mnras_515_2022}.
The magneto-hydrodynamical simulations are conducted in a 2.5-dimensional, axisymmetric cylindrical coordinate system. The simulation box extends over the range $[O;R_{\rm max}]\times[z_{\rm min}; z_{\rm max}]$ and is discretized using a uniform grid of $N_{\rm R}\times N_{\rm z}$ cells. 
Consequently, the spatial resolution is consistent along both directions, with each grid cell having a size of $\Delta = R_{\rm max}/N_{\rm R}$.
We employ two different spatial resolutions throughout the evolutionary process. 
During the progenitor star wind phases, the circumstellar medium is resolved using a grid resolution of $N_{\rm R}=2000$ and $N_{\rm z}=4000$ cells. 
The stellar wind is implemented as an internal boundary condition within a sphere centred at the origin of the computational domain, with a radius of $20\Delta$, following the standard procedure outlined in \citet{comeron_aa_338_1998}.

At the immediate pre-supernova stage, we remap the solution for the circumstellar medium onto a finer grid with $N_{\rm R}=3000$ and $N_{\rm z}=6000$ cells. 
The supernova ejecta is confined within a central sphere of radius $r_{\rm max}$, as described in section~\ref{initial_conditions}. 
Simultaneously, the pulsar wind is imposed within a sphere of radius $r_{\rm ns\,wind}=20\Delta$, also detailed in section~\ref{initial_conditions}.
Due to our choice of an asymmetric coordinate system, we are compelled to align the directions of the pulsar spin axis and the symmetry axis of the computational domain to be the same.


In this paper, we study the evolution of the circumstellar medium influenced by the magnetized wind emitted by a massive star with a mass of $35\, \rm M_{\odot}$ in two distinct types of external medium: the magnetized and unmagnetized warm phases of the Galactic plane in the Milky Way. 
We refer to these models as \normalfont{Run-35-HD-0.79-PWN} and \normalfont{Run-35-MHD-0.79-PWN}. 
In the magnetised external medium case, the adopted strength of the background magnetic field 
is set to that measured in the spiral arms of the Galaxy, with an average strength of 
$B_{\rm ISM}=7\, \mu \rm G$ \citep[see][]{Drain_book_2011piim.book.....D}.
%
The main parameters utilized in the two cases investigated in this paper are provided in Table \ref{tab:table1}. 
For a more comprehensive description of the model and the implemented strategy, please refer to \citet{Meyer_etal_2023MNRAS.521.5354M} and \citet{meliani_mnras_515_2022}, where detailed explanations can be found.


\begin{figure*}
        \centering
        \includegraphics[width=0.85\textwidth]{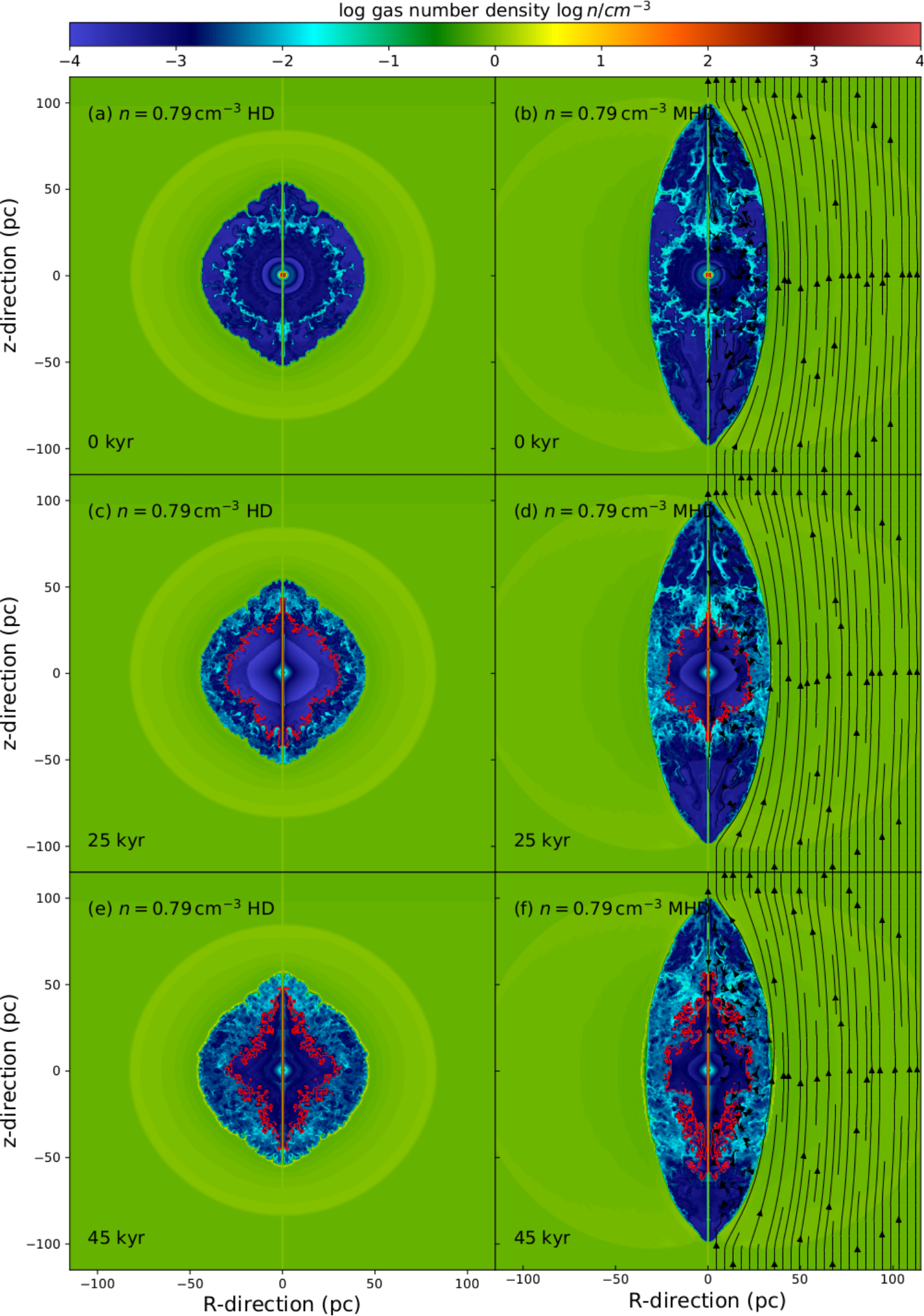}  \\
        \caption{
        Number density fields in our magneto-hydrodynamical simulation of 
        the pulsar wind nebula forming in the supernova remnant of a static 
        $35\, \rm M_{\odot}$ star rotating with $\Omega_{\star}/\Omega_{\rm K}=0.1$ 
        in an unmagnetised (left) and magnetised (right) ISM. 
        The red contours highlight the region with a $50\%$ contribution of 
        pulsar wind material,  i.e. the contact discontinuity. 
        The streamlines in the right-hand side of panels b,d,f mark the 
        ISM magnetic field lines. 
        }
        \label{fig:SNR_time_evolution}  
\end{figure*}

The numerical simulations are conducted using the \textsc{pluto} code \citep{mignone_apj_170_2007,migmone_apjs_198_2012,vaidya_apj_865_2018}\footnote{http://plutocode.ph.unito.it/}
and we solve the following set of equations, 
\begin{equation}
	   \frac{\partial \rho}{\partial t}  + 
	   \bmath{\nabla}  \cdot \big(\rho\bmath{v}) =   0,
\label{eq:mhdeq_1}
\end{equation}
\begin{equation}
	   \frac{\partial \bmath{m} }{\partial t}  + 
           \bmath{\nabla} \cdot \Big( \bmath{m} \otimes \bmath{v}  
           - \bmath{B} \otimes \bmath{B} + \bmath{\hat I}p_{\rm t} \Big)  \bmath{0},
\end{equation}
\begin{equation}
	  \frac{\partial E }{\partial t}   + 
	  \bmath{\nabla} \cdot \Big( (E+p_{\rm t})\bmath{v}-\bmath{B}(\bmath{v}\cdot\bmath{B}) \Big)  
	  = \Phi(T,\rho),
\label{eq:mhdeq_3}
\end{equation}
and
\begin{equation}
	  \frac{\partial \bmath{B} }{\partial t}   + 
	  \bmath{\nabla} \cdot \Big( \bmath{v} \otimes \bmath{B} - \bmath{B} \otimes \bmath{v} \Big)  =
	  \bmath{0},
\label{eq:mhdeq_4}
\end{equation}
with the gas density $\rho$, velocity $v$, momentum $\bmath{m}=\rho\bmath{v}$ and 
magnetic field $\bmath{B}$, as well as the the total pressure $p_{t}$ and the 
energy of the gas
\begin{equation}
	E = \frac{p}{(\gamma - 1)} + \frac{ \bmath{m} \cdot \bmath{m} }{2\rho} 
	    + \frac{ \bmath{B} \cdot \bmath{B} }{2}. 
\label{eq:energy}
\end{equation}
The sound speed of the medium reads 
\begin{equation}
c_{\rm s} = \sqrt{ \frac{ \gamma p }{ \rho } } ,
\end{equation}
where the adiabatic index is $\gamma=5/3$. 
Last, radiative cooling by optically-thin 
processes and photo-heating are included into the equations via the term $\Phi(T,\rho)$, 
with the gas temperature $T$, accounting for the prescriptions of~\citet{meyer_2014bb}. 
Regarding the cooling/heating processes of the gas, we assume the gas to be optically thin throughout the entire progenitor's life. After this point, with the launch of the pulsar wind, the cooling and heating 
mechanisms are disabled. 
We employ a Godunov-type numerical scheme with the Harten-Lax-van Leer approximate Riemann solver (HLL) and utilize the 8-waves magnetic field formulation \citep{Powell1997}. 
For time integration, a third-order Runge-Kutta scheme is employed, with the time step controlled by the Courant-Friedrichs-Lewy (CFL) number.

The numerical simulations are performed at the North-German Supercomputing Alliance (HLRN\footnote{https://www.hlrn.de/}) using the {\sc LISE} cluster in Berlin, which is equipped with Cray XC40/30 processors.

\section{Results}
\label{sect:results}

In this section, we will analyze the results of the evolution of the PWN within the supernova remnant and circumstellar medium of the progenitor star in both the unmagnetized and magnetized cases. 
Our focus will be on investigating the influence of the magnetic field of the progenitor star and the external medium on the shape and dynamics of the PWN.

\subsection{Model with unmagnetised ISM}
\label{results_B}

In Figure~\ref{fig:SNR_time_evolution}, the density contour is shown for the unmagnetized case \normalfont{Run-35-HD-0.79-PWN} (left panels) and the magnetized case \normalfont{Run-35-MHD-0.79-PWN} (right panels) at different evolution times, from to top to the bottom. 
The density contour is represented in logarithmic scale in $\rm cm^{-3}$ units. In both cases, the red contour marks indicate the region of the plerionic supernova remnants where the contribution of the pulsar wind reaches $50$ times of the number density.

In Figure~\ref{fig:SNR_time_evolution}a (top-left), we present the pre-supernova circumstellar medium. At this stage, it forms a large-scale quasi-spherical stellar bubble~\citep{weaver_apj_218_1977}, and its spherical forward shock extends to distances of approximately $90\, \rm pc$. Throughout the star's evolution, the stellar wind interacts strongly with the ambient medium. Each phase of evolution contributes to the formation of successive shock structures, which appear in order from the farthest to the nearest region to the star. The thick and dense shell located farthest from the star, with a radial extent of $\geq 50\, \rm pc$, is the result of the interaction between the stellar wind and the ISM, and it occurs mainly during the main-sequence phase~\citep{freyer_apj_594_2003,dwarkadas_apj_630_2005,freyer_apj_638_2006,dwarkadas_apj_667_2007}. 
In the central region, within a radius of less than $20\, \rm pc$, a low-density cavity is formed due to the continuous outflow of the free stellar wind during the Wolf-Rayet phase. 
This cavity is surrounded by successive dense shells resulting from the interactions between the Wolf-Rayet wind and the slower wind from the preceding red-giant phase. 
The first shell, extending to approximately $35\, \rm pc$, is dense and exhibits unstable behaviour. 
Subsequently, a second, less dense shell is formed due to the interaction between the red-giant wind and the main-sequence wind. 
Additionally, the main-sequence wind interacts with the surrounding ambient medium, forming an external dense shell that is limited by the contact discontinuity surface.

It is worth noting that the contact discontinuity, which marks the interface between the wind and the ISM, exhibits a slightly aspherical morphology, particularly in the region close to the symmetry axis. 
This aspherical shape is influenced by the presence of the magnetic field and the star's rotation. 
The variations between the bubbles depicted in Fig.~\ref{fig:SNR_time_evolution}a and Fig.~1 of \citet{meyer_515_mnras_2022} highlight this effect. 
Furthermore, the grid's proximity to the near-polar axis amplifies this asymmetry.
Moving on to Fig.~\ref{fig:SNR_time_evolution}c (middle-left), we can observe the supernova remnant at $25\, \rm kyr$ after the explosion. 
The expanding shock wave from the supernova remnant propagates outward, sweeping up and pushing away all the previously formed dense shells associated with the successive stellar winds. 
 As the shock wave reaches the contact discontinuity surface between the 
main-sequence stellar wind and the ISM, it interacts with this surface, causing 
reflection, 
as described in  
\citet{meyer_mnras_450_2015,meyer_mnras_502_2021,meliani_mnras_515_2022}. 
This interaction and reverberation contribute to the observed structure and morphology of the supernova remnant. 


 
After the supernova explosion, a pulsar wind with high initial mechanical 
luminosity, $\dot{E}_{\rm psr, 0}=10^{38} \rm erg  s^{-1}$, is launched. 
However, this luminosity decreases over time according to Eq. \ref{eq:energy} 
of \citet{Pacini_1967Natur.216..567P}. 
This pulsar wind interacts with the dense supernova ejecta~\citep{swaluw_aa_420_2004}, resulting in the formation of a complex structure as described in \citet{meliani_mnras_515_2022} for a runaway progenitor star with a zero-age main-mass of $20\, \rm M_{\odot}$.
%
%
%
%
Within this structure, the central region of the plerion is occupied by the freely-expanding pulsar wind. Surrounding the central region, a shell of shocked pulsar wind is formed, resulting from the interaction of the pulsar wind with the expanding supernova remnant. A pulsar wind termination shock is formed at the interface between the unperturbed pulsar wind and the shocked  pulsar wind.
%
%
The outermost region of the pulsar wind nebula behind the termination shock contains the contact discontinuity. This contact discontinuity marks the interface between the supernova ejecta and the shocked pulsar wind (depicted by the red contour in Fig.~\ref{fig:SNR_time_evolution}). 
Beyond the contact discontinuity, a transmitted pulsar wind forward shock propagates through the still 
unshocked supernova ejecta and further travels into the surrounding medium.

The pulsar wind contact discontinuity undergoes expansion to larger radii due to the fast rotation of the magnetized neutron star. 
%
%
This expansion leads to the characteristic shape with an equatorial torus and an elongated polar jet, as found by \citet{komissarov_mnras_349_2004, Del_Zanna_etal_2006A&A...453..621D,Porth_etal_2014MNRAS.438..278P, Olmi_etla_2016JPlPh..82f6301O}. However, it's important to note that due to limitations in the numerical scheme applied to the 2D symmetry axis, the jet along the polar axis may appear more elongated than it would in a full 3D simulation. 
Nevertheless, despite these limitations, the general behaviors of the PWN remains accurate.
This shape can be observed at a later time, specifically $45\, \rm kyr$ after the explosion, as shown in Fig.~\ref{fig:SNR_time_evolution}e.
As the contact discontinuity surface expands, it encounters Rayleigh-Taylor instabilities due to the significant differences in density and velocity between the pulsar wind and the supernova ejecta. These instabilities are further amplified by the reverberation of the reverse shock from the supernova ejecta, as illustrated in Fig.~\ref{fig:SNR_time_evolution}e.

\subsection{Model with magnetized ISM}
\label{results_ISM}

During the main-sequence phase of a massive star, the influence of the ISM magnetic field becomes particularly significant. 
During this phase, the interaction between the stellar wind and the magnetized ISM carves out a large-scale circumstellar wind bubble. 
This wind bubble plays a crucial role in shaping the propagation of the supernova forward shock. 
Additionally, the wind bubble's presence influences the pulsar wind's dynamics, further highlighting the interplay between the stellar wind, the ISM magnetic field, and the subsequent evolution of the system.  
We will describe it in detail in the following. 
In Fig.~\ref{fig:SNR_time_evolution}b (top-right), we can observe the circumstellar medium surrounding the massive star in the presence of a magnetized ISM, as represented in the model \normalfont{Run-35-MHD-0.79-PWN}. 
The black arrows indicate the magnetic field lines of the ISM, which are initially aligned with the polar axis.
The overall structure of the circumstellar medium in the presence of the magnetized ISM remains similar to the unmagnetized model (\normalfont{Run-35-HD-0.79-PWN}, Fig.~\ref{fig:SNR_time_evolution}a). 
However, the morphology of the shocked shells within the low-density cavity, up to the contact discontinuity between the shocked stellar wind and the shocked ISM, appears to be more elongated along the polar axis due to the influence of the ISM magnetic field.

Indeed, as the expanding stellar bubble interacts with the magnetized ISM, it compresses the magnetic field lines, increasing magnetic pressure and tension along the polar axis. This phenomenon has been extensively studied and described in detail in \citet{vanmarle_584_aa_2015}.
During the last evolution phase, when the Wolf-Rayet wind material reaches the main-sequence termination shock, it undergoes reflection near the equator. This anisotropic reflection causes a change in the direction of propagation of the shocked material, resulting in the loss of the initially spherical shape of the shocked shell from the Wolf-Rayet wind. 
The interaction with the magnetized ISM further influences the shape and dynamics of the shocked shell, leading to the observed rectangular morphology of the resulting supernova ejecta.
Furthermore, as the expanding supernova blast wave propagates within the elongated cavity (as shown in the left panel of Fig.~\ref{fig:SNR_time_evolution}), it interacts with the reflected dense shells resulting from the Wolf-Rayet wind and the elongated contact discontinuity. These interactions lead to anisotropic reverberation at the contact discontinuity of the supernova ejecta. As a result, the shape of the supernova ejecta becomes rectangular, reflecting the influence of the asymmetric interactions with the elongated structure induced by the magnetized circumstellar medium. This mechanism is specifically described within the context of the remnant Puppis A in \citet{meyer_515_mnras_2022}.

In Fig.~\ref{fig:SNR_time_evolution}d and f, the influence of the magnetized ISM on the shaping of the PWN can be observed. The ISM magnetic field, which plays a significant role in determining the morphology of the circumstellar medium and supernova blastwave, also affects the confinement and shape of the pulsar wind.
Under the influence of the ISM magnetic field, the reflected and the supernova blastwave adopts a rectangular morphology along the direction perpendicular to the magnetic field. This happens because the ram pressure of the supernova ejecta is directed towards the polar axis, causing compression and confinement of the pulsar wind in that direction. 
In contrast, in the direction parallel to the magnetic field, the pressure exerted by the supernova ejecta is lower, resulting in a more extended shape of the PWN.
This interplay between the magnetic field of the ISM, the reverse shock, and the pulsar wind contributes to the complex and asymmetric morphology observed in the PWN, as depicted in Fig.~\ref{fig:SNR_time_evolution}d and f.

Indeed, the presence of a magnetized ISM influences the propagation of the PWN, resulting in distinct behavior compared to an unmagnetized ISM. In the magnetized ISM model (\normalfont{Run-35-MHD-0.79-PWN}), the expansion of the PWN is less pronounced in the equatorial plane compared to the hydrodynamical simulation (\normalfont{Run-35-HD-0.79-PWN}), as illustrated in Fig.\ref{fig:SNR_time_evolution}c and d. 
As time progresses, at a later evolution time of $45\, \rm kyr$ as depicted in Fig.\ref{fig:SNR_time_evolution}f, the pulsar wind continues to be channeled along the direction of the ISM's magnetic field, leading to the formation of a stretched PWN.
The presence of the ISM magnetic field affects the dynamics of the PWN and leads to enhanced instabilities at the termination shock of the pulsar wind. 
These instabilities, which arise from the interaction between the pulsar wind and the magnetized ISM, are more pronounced in the magnetized ISM model (\normalfont{Run-35-MHD-0.79-PWN}) compared to the hydrodynamical simulation (\normalfont{Run-35-HD-0.79-PWN}).

Our models provide compelling evidence that the morphology of the PWN inside a subsequent supernova, when the progenitor  massive static star is located in the Galactic plane, is strongly influenced by the distribution of the magnetic field in the ambient medium. 
The contrasting evolution and instabilities observed in the magnetized and unmagnetized cases emphasize the significant role played by the interstellar medium's magnetic field in shaping the dynamics and morphology of the PWN. 
These findings underscore the importance of considering the magnetic field effects when studying the evolution of PWN and their interaction with the surrounding environment.


\begin{figure*}
        \centering
        \includegraphics[width=0.985\textwidth]{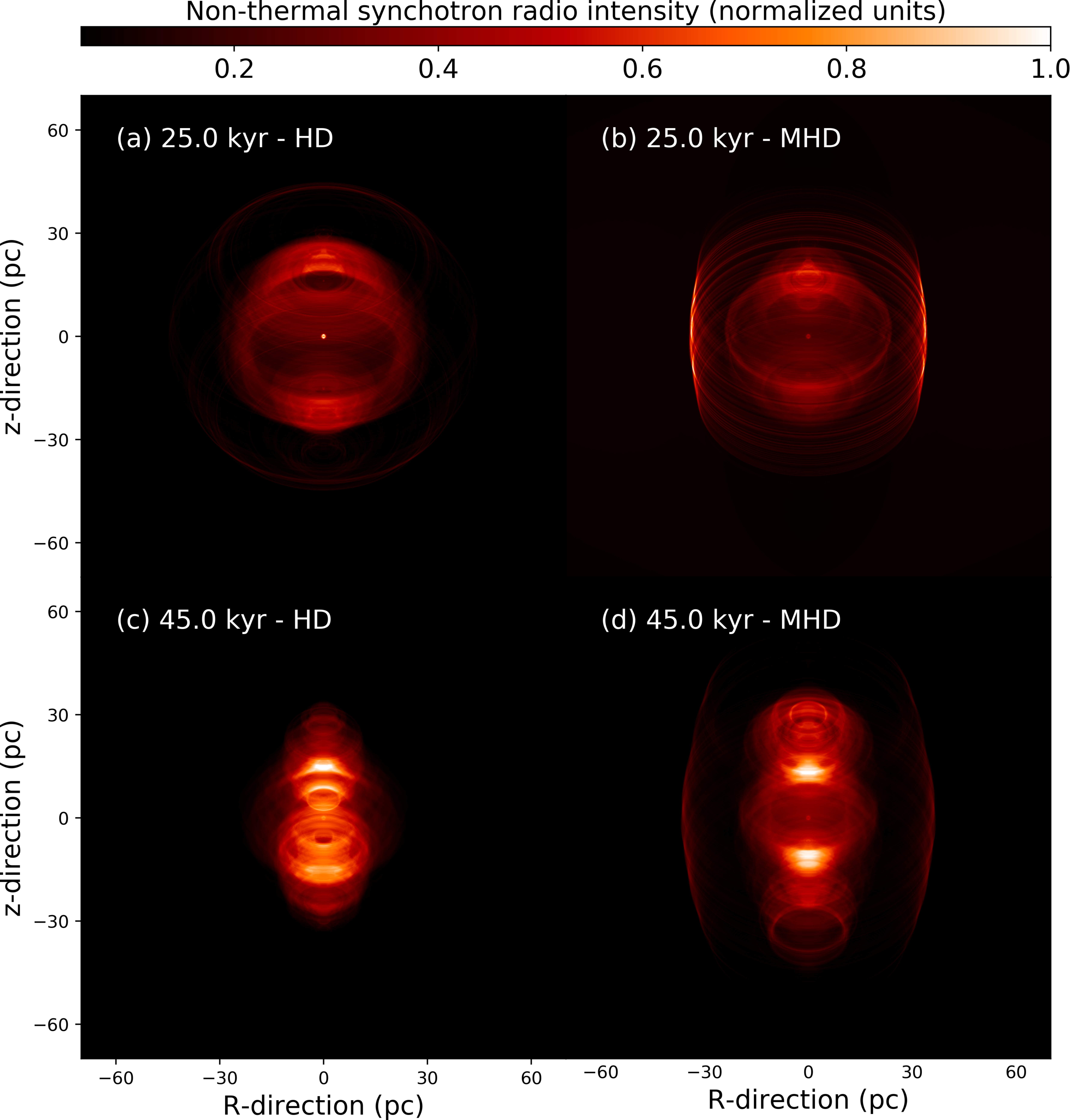}  \\
        \caption{
        Normalised radio synchrotron emission map of the plerionic supernova 
        remnants with an inclination angle of $\theta_{\rm obs} = 45\degree$ 
        between the observer’s line of sight and the nebula's symmetry axis. 
        The left-hand panels correspond to the hydrodynamical model (\normalfont{Run-35-HD-0.79-PWN}), 
        and right-hand panel to the model with magnetised ISM (\normalfont{Run-35-MHD-0.79-PWN}).
        The top figures show the remnants at time $25\, \rm kyr$ and the bottom 
        figures display them at time $45\, \rm kyr$. 
        }
        \label{fig:emission_maps}  
\end{figure*}

\section{Discussion}
\label{sect:discussion}

In this section, we will discuss the applications and limitations, 
of our model. We will also examine the non-thermal characteristics of the simulated pulsar wind nebulae and compare our findings to existing observational data. 
By doing so, we aim to provide a comprehensive analysis of our model's strengths and weaknesses and assess its compatibility with the observed properties of pulsar wind nebulae.

\subsection{Model limitations}
\label{model_limitation}

Let us first consider four aspects central to the model.
First, the simulations conducted in this study are two-dimensional, assuming axisymmetry and not accounting for variations in the supernova progenitor or the pulsar's spin. While this approach offers computational efficiency and valuable insights, it's essential to acknowledge that a fully three-dimensional treatment is not only important to capture the realistic properties of the ISM but also crucial for a comprehensive understanding of the pulsar wind nebula and the supernova remnant. A 3D model would better represent the complex interactions of the PWN and 
supernova remnant with the surrounding medium, including the realistic behaviour of magnetic fields.
Moreover, the magnetization of the pulsar wind is a fundamental parameter that plays a significant role in the evolution of the PWN and supernova remnant. While we have considered a weak magnetization of the pulsar wind in this study, it's essential to discuss its implications thoroughly. State-of-the-art simulations in both 2D and 3D have shown that the strength and longitudinal variation of magnetization are subjects of debate \citep{Coroniti_2017ApJ...850..184C,Olmi_etla_2016JPlPh..82f6301O}. Future investigations will explore the influence of higher magnetization on the evolution of the PWN in its interaction with supernova remnant and circumstellar medium.

Furthermore, we acknowledge that our modelling of the pulsar wind involves simplified assumptions. A more realistic modelling approach should also involve a better physical description of the wind properties, including its relativistic speed and composition. Addressing these aspects will be crucial in future research for a more comprehensive understanding of the system's dynamics and morphology.
Another aspect to consider is the absence of pulsar motion in the simulations. Incorporating pulsar motion would introduce additional complexities and offer a more realistic representation of the interaction between the pulsar wind and the surrounding medium. Furthermore, accounting for the oblique rotation of the pulsar's magnetic axis would allow for a more accurate reproduction of the observed characteristics of the PWN.
These are important considerations for future research. The chosen two-dimensional setup and static pulsar position provide valuable insights into general behaviour and trends. However, future investigations can explore the impact of three-dimensional effects, pulsar motion, higher magnetization, and improved modelling of the pulsar wind to obtain a more comprehensive characterization of the system's dynamics and morphology.

\subsection{Non-thermal emission}
\label{maps}

To enhance the comparison between our MHD models of the PWN embedded in an elongated circumstellar medium and the available observational data, we performed radiative transfer calculations to generate synthetic images that accurately capture the non-thermal emissions, particularly synchrotron emissions in the radio band. 
These calculations were specifically carried out at the different evolution stages of the PWN that were previously discussed.
The synchrotron radio emission was calculated by considering a non-thermal electron spectrum described by the expression, 
\begin{equation}
N(E) = K E^{-s} \propto n E^{-2},
\label{eq:N}
\end{equation}
where here, $n$ represents the gas number density, $s$ is the spectral index and $E$ 
denotes the energy of the non-thermal electrons in the post-shock region of the advancing blast wave. 
The emission coefficient is given by:
\begin{equation}
j_{\rm sync}(\nu) \propto n B_{\perp}^{ (s+1)/2 } \nu^{ -(s-1)/2 },
\label{eq:coeff}
\end{equation}
being $\nu$ the observed frequency and $B_{\perp}$ the component of the magnetic field perpendicular to the observer's line-of-sight.

Intensity maps were obtained by performing the projection given by, 
\begin{equation}
I = \int_{\rm SNR} j_{\rm sync}\left(\theta_{\rm obs}\right) dl, 
\label{eq:intensity}
\end{equation}
where $\theta_{\rm obs}$ denote the inclination angle of the remnant with respect to the sky plane.
These calculations were conducted using the radiative transfer code RADMC-3D\footnote{https://www.ita.uni-heidelberg.de/dullemond/software/radmc-3d/}, and the methodology described in detail by \citet{meyer_515_mnras_2022}.
%
Note that since the investigated numerical simulations are non-relativistic, they do not account for the beaming effect, and this issue will be addressed in our upcoming work.

Figure \ref{fig:emission_maps} illustrates the normalized emission maps representing our numerical simulations, specifically \normalfont{Run-35-HD-0.79-PWN} (left-hand panels) and \normalfont{Run-35-MHD-0.79-PWN} (right-hand panels), showcasing the non-thermal synchrotron emissions in the radio waveband. The top panels correspond to $25 \rm kyr$, while the bottom panels depict a time of $45 \, \rm kyr$. The intensity is plotted assuming an observer angle ($\theta_{\rm obs}$) of $45\degree$, representing the angle between the plane of the sky and the plane of symmetry of the supernova remnant.
Figure \ref{fig:emission_maps}a displays the pulsar wind nebula at the age of $25\, \rm kyr$ within an unmagnetized ISM. 
As highlighted in~\citet{meliani_mnras_515_2022}, no trace of the circumstellar medium is visible in the emission maps because of the absence of the ISM magnetic field. Indeed, the emission map focuses on the pulsar wind and its associated nebula. 
The image reveals an ovoidal shape, with slightly brighter regions observed at the polar and dimmer regions in the equatorial plane. 
This brightness variation can be attributed to the toroidal component of the pulsar wind, which applies lateral pressure on the pulsar wind material, causing it to be displaced sideways in the equatorial plane.

At a later evolution time, with a pulsar age of $45\, \rm kyr$, the radio synchrotron map of the PWN in an unmagnetized ISM is shown in Fig. \ref{fig:emission_maps}c. 
The PWN exhibits a jet-torus-like shape, with brighter regions observed at the polar zones. 
These bright regions result from the strong interaction between the pulsar wind and the supernova ejecta along the pulsar's rotational axis. 
On the other hand, in the equatorial plane, the strong pulsar wind, driven by the centrifugal force and toroidal magnetic field pressure~\citep{komissarov_mnras_349_2004}, extends outward. The gas is more diluted in this region, which explains why the equatorial plane is not the brightest region in the hydrodynamical plerion model \normalfont{Run-35-HD-0.79-PWN}. 
In the case of a magnetized ISM, significant changes are observed in the synthetic radio image. 
The corresponding image is shown at $25\, \rm kyr$ in Fig. \ref{fig:emission_maps}b. 
It reveals the presence of two bright arcs parallel to the direction of the ISM magnetic field. 
These arcs, observed in our axisymmetric setup and aligned with the pulsar's rotation axis, are formed as a result of the interaction between the supernova ejecta and the contact discontinuity between the stellar wind and the magnetized ISM within the elongated cavity~\citep{meyer_515_mnras_2022}. 
The influence of the ISM magnetic field plays a crucial role in shaping these arcs, ultimately forming a PWN enclosed within a rectangular supernova remnant.

 Fig. \ref{fig:emission_maps}d depicts the older remnant within a magnetized ambient medium, showcasing characteristics of both a supernova shock wave that has interacted with the cavity's border and the growing pulsar wind nebula inside it. 
 The presence of the pulsar wind prevents the reverberation of the supernova shock wave towards the centre of the explosion, as described in~\citet{meyer_515_mnras_2022}, resulting in an empty region near the rotating neutron star. 
 The overall morphology of the plerionic remnant still exhibits features of a rectangularly reflected supernova shock wave, with the pulsar wind distributed as an elongated structure. 
 The brightest regions are observed as two polar spots located beyond the termination shock of the pulsar wind.

\subsection{Comparison with observations}
\label{comparison}


The models presented in this study focus on the evolution of the circumstellar medium surrounding static high-mass stellar objects that eventually undergo supernova explosions, leaving behind a static pulsar. We aim to investigate the formation of elongated pulsar wind nebulae, similar to those observed in~\citet{igoshev_mnras_494_2020}. It is important to note that these elongated PWN, where the leptonic wind is channelled into the cavity created by the stellar wind shaped by the organized ISM magnetic field, should not be confused with the long tails observed behind the bow shocks of runaway pulsars
\citep[e.g.,][]{Bucciantini_aa_387_2002, bucciantini_mnras_478_2018,luca_apj_765_2013,barkov_mnras_484_2019}.
The class of torus/jet-like pulsar wind nebulae, as classified in the catalogue based on {\sc Chandra} X-ray data, provides strong support for the conclusions drawn from our model. These objects naturally exhibit both an equatorial structure and a jet/counter-jet system, as observed in studies such as~\citet{kargaltsev_aipc_1248_2010,kargaltsev_apj_745_2012} and references therein. Notable examples include the famous Crab nebula with its twisted double jet~\citep{Mignone_mnras_436_2013} and the Vela supernova remnant. Magneto-hydrodynamical models have successfully reproduced such structures without considering the stellar wind or supernova ejecta as initial conditions, as demonstrated in~\citet{klingler_aas_2014}.

The influence of the environment on the morphology of pulsar wind tails/jets has been demonstrated in cases such as the Geminga pulsar wind nebula, which exhibits two curved antennae representing its jets/counter-jet that bend under the influence of the bow shock formed due to the interaction between the fast pulsar motion and the surrounding medium~\citep{posselt_apj_835_2017}. Similar effects have been observed in the case of B0355+54~\citep{klingler_aas_2014}. We propose that the pre-supernova environment plays a similar role, and further modelling efforts are highly desirable, as discussed in~\citet{meliani_mnras_515_2022}.
The peculiar morphology of certain pulsar wind nebulae, which cannot be classified as either torus/jet-like objects or bow shock/tail systems, may result from their interaction with a particularly complex surrounding medium. This medium could be shaped by the asymmetric stellar wind during the evolved phases of the progenitor's pre-supernova life, which influences the forward shock of the ejecta and causes aspherical propagation \citep{velazquez_mnras_519_2023,villagran_mras_2023}.

\section{Conclusion}
\label{sect:conclusion}

This paper presents a study on the modelling of PWN  in core-collapse supernova remnants associated with static massive stars in the warm phase of a magnetized spiral arm of the Milky Way.
By utilizing 2.5-dimensional simulations, we demonstrate that the reflection of the supernova blast wave against the elongated contact discontinuity between the stellar wind and magnetise ISM of the magnetically elongated stellar wind cavity in the progenitor's circumstellar medium has a significant impact on the morphology of the resulting PWN. 
This phenomenon might be responsible for forming rectangular supernova remnants, such as Puppis A, as described in \citet{meyer_515_mnras_2022}.
The reverberation of the shock wave leads to the compression of the pulsar wind and imposes a preferred expansion direction perpendicular to the plane of the pulsar's spin. 
As a result, the PWN within the rectangular supernova remnant becomes elongated rather than adopting the jet-torus-like shape typically observed in previous studies, as described by \citet{komissarov_mnras_349_2004}.

The radio synchrotron emission maps of plerionic supernova remnants exhibit a complex morphology that evolves over time. 
Initially, the morphology is characterized by a young, growing, ovoidal PWN combined with the rectangular shape produced by the interaction between the supernova ejecta and the walls of the unshocked stellar wind cavity of the progenitor star. 
This interaction gives rise to the rectangular appearance observed in Puppis A, as discussed in \citet{meyer_515_mnras_2022}.
As time progresses, the influence of the ISM magnetic field becomes more prominent in shaping the remnant's morphology. 
The channelling effect of the pulsar wind into the elongated circumstellar wind cavity of the progenitor extends along the pulsar's rotation axis. Instabilities at the interface between the pulsar wind and the ejecta result in a knotty nebula, manifesting as bright spots within the plerion.
The irregular shapes observed in many pulsar wind nebulae may indicate the complex nature of the surrounding environment, influenced by both the distribution of material in the ambient medium and the stellar wind history of the supernova progenitor. 
In this complex environment, the interaction between the supernova ejecta and the pulsar wind gives rise to observed irregular morphologies.

%


\section*{Acknowledgements}

The authors acknowledge the referee for numerous constructive comments. 
DMAM thanks O. Petruk for his comments on the manuscript. 
The authors acknowledge the North-German Supercomputing Alliance (HLRN) for providing HPC 
resources that have contributed to the research results reported in this paper. PFV acknowledges 
financial support from the PAPIIT IG100422 grant. 
DFT has been supported by grants PID2021-124581OB-I00 funded by MCIN/AEI/10.13039/501100011033, 2021SGR00426 of the Generalitat de Catalunya, by the program Unidad de Excelencia María de Maeztu CEX2020-001058-M, and by 
MCIN with funding from European Union NextGeneration EU (PRTR-C17.I1).

\section*{Data availability}

This research made use of the {\sc pluto} code developed at the University of Torino  
by A.~Mignone (http://plutocode.ph.unito.it/). 
and of the {\sc radmc-3d} code developed at the University of Heidelberg by C.~Dullemond 
(https://www.ita.uni-heidelberg.de/$\sim$dullemond/software/radmc-3d/).
The figures have been produced using the Matplotlib plotting library for the 
Python programming language (https://matplotlib.org/). 
The data underlying this article will be shared on reasonable request to the 
corresponding author.


\bibliographystyle{mnras}
\bibliography{grid} 

\bsp	
\label{lastpage}
\end{document}